\documentclass[12pt]{article}
\def\vec#1{\mbox{\boldmath $#1$}}

\begin{document}

\noindent
{\Large
\bf
A new representation of the many body wave function
and its application as a post Hartree-Fock 
energy variation method}\\
\begin{center}
Wataru Uemura\\
Department of Physics, University of Tokyo\\
February 24, 2011
\end{center}

\section{Introduction}

In this paper, we introduce a new representation
of many body electron wave function
and a few calculation results of the ground state energies of many body systems
using that representation,
which is systematically better than the Hartree-Fock approximation.\par
The fundamental principle of 
condensed matter physics and chemistry
is given in the many body schrodinger equation,
which is
\begin{equation}
H = \sum _{i=1}^N(-(1/2)\nabla _i^2+v(\vec{r}_i)) + \sum _{(i, j)}1/|\vec{r}_i-\vec{r}_j|,
\end{equation}

\begin{equation}
H\psi (x_1\cdots x_N) = E\psi (x_1\cdots x_N).\label{schrodinger}
\end{equation}
This is a schrodinger equation of $N$ electrons 
in the external potential $v(\vec{r})$ from nuclei
under the Born-Oppenheimer approximation.
The electron wave function $\psi (x_1\cdots x_N)$
must be antisymmetric in the exchange of 
arbitrary two spin coordinates $x_i$ and $x_j$.
One of the central problem in condensed matter physics and chemistry
is to find the solution of (\ref{schrodinger})
and the ground state energy $E_0$ of the given system.
There are many attempts to solve this problem.
In this section, we briefly look over
the Hartree-Fock approximation (HF)
and the Density functional theory (DFT) or 
Local density approximation (LDA).\par
In the Hatree-Fock approximation,
the many body wavefunction $\psi (x_1\cdots x_N)$
is approximated by a single slater determinant $\psi _{SL}$,
and this gives limitation to the accuracy of 
calculated ground state energy $E_{HF}$.
The reason is that an antisymmetric wave function $\psi (x_1\cdots x_N)$
is not always given as a single slater determinant $\psi _{SL}$.
Rather, $\psi (x_1\cdots x_N)$ can be expanded in 
a linear combination of $_MC_N$ slater determinants 
in the space of given orbital set $\psi _1,\cdots,\psi _M$.
In this way, one can rearrange the many body electron problem
into the diagonalization of $_MC_N \times _MC_N$ matrix.
This method is known as Full CI method.
Full CI always gives the exact ground state energy
in the space of a given orbital set $\psi _1,\cdots,\psi _M$.
However, when the number of the electrons $N$ and 
the orbitals $M$ increases 
as much as a few dozens, 
the dimension of the matrix $_MC_N$ increases exponentially with $N$ and $M$
and Full CI calculation becomes practically impossible.\par
In the Hartree-Fock approximation, the variables of the variation
are the $N$ orbitals $\psi _1,\cdots,\psi _N$.
In contrast, in the density functional theory, 
the variable is the one electron density $\rho (\vec{r})$.
This $\rho (\vec{r})$ can be uniquely deduced from the 
given antisymmetric wave function $\psi (x_1\cdots x_N)$.
The essence of the DFT is that 
the expectation value of the sum of the kinetic energy $T$
and the coulomb repulsion energy $U$
in the ground state can be given by
a unique functional of the one electron density, $F[\rho (\vec{r})]$.
However, the exact form of this functional $F[\rho (\vec{r})]$
is not known until today.
Therefore, in LDA calculation, 
several approximated form of this functional 
is used and they are not exact.
In LDA, the $v$-representability of $\rho (\vec{r})$ is generally assumed,
if one uses integers for the occupation numbers.
In this case, the calculated ground state energy $E_{LDA}$
and the one electron density $\rho (\vec{r})$
are derived from a non interacting single slater determinant $\psi _{SL}$.
This may be the reason why the LDA calculation
does not work well 
in the so called strongly correlated systems.\par
In the next section, we introduce a new representation
of the antisymmetric wave function
which is an extension of the slater determinant and therefore
not always non interacting.

\section{A new representation of many body wave function}

In the many body schrodinger equation (\ref{schrodinger}),
the variable is the $N$ body antisymmetric wave function $\psi (x_1\cdots x_N)$.
However, this function is apparently not suited for any variational calculation
because if one takes $m$ spatial grids for one variable $x_i$,
then the total grids of the function is proportional to $m^N$
and becomes progressively impossible to stock in the memory of the computer
when he increases $N$ to only a few dozens.
One way to express this wave function $\psi (x_1\cdots x_N)$
is to expand it in a given $M$ orbital set $\psi _1,\cdots,\psi _M$.
In this way, the wave function is expressed as
\begin{equation}
\psi (x_1\cdots x_N) = \sum _{i_1\cdots i_N} A_{i_1\cdots i_N}\psi _{i_1}(x_1)\cdots \psi _{i_N}(x_N).
\label{tensor}
\end{equation}
In order to keep $\psi (x_1\cdots x_N)$ antisymmetric,
the rank-$N$ and dimension-$M$ tensor $A_{i_1\cdots i_N}$
should be antisymmetric tensor.
This antisymmetric tensor has $_MC_N$ degree of freedom
and still be hard to take variation except for very small systems.\par
Here we introduce a new representation of this antisymmetric tensor $A_{i_1\cdots i_N}$:
\begin{equation}
A_{i_1\cdots i_N} = \sum _{i=1}^{K} c_i c_{i_1}^i \cdots c_{i_N}^i 
\epsilon _{i_1i_2}\epsilon _{i_1i_3} \cdots \epsilon _{i_{N-1}i_N}.
\label{expansion}
\end{equation}
In this representation, each $c_{i}^1,\cdots,c_{i}^K$
represents a vector of dimension $M$.
For this reason, let us tentatively call this representation as the vector product.
$c_1,\cdots,c_K$ are arbitrary coefficients.
$\epsilon _{ij}$ is a rank-$2$ and dimension-$M$ tensor which is defined by
\begin{eqnarray}
\epsilon _{ij} = \left\{ \begin{array}{ll}
1 & (i<j), \\
-1 & (i>j), \\
0 & (i=j). \\
\end{array} \right.
\end{eqnarray}\par
It is easy to verify that $\epsilon _{i_1\cdots i_N} \equiv \epsilon _{i_1i_2} \cdots \epsilon _{i_{N-1}i_N}$, 
$N(N-1)/2$ product of $\epsilon _{ij}$,
is rank-$N$ and dimension-$M$ antisymmetric tensor
which takes the value of only $1$, $-1$ or $0$ for any indices $i_1\cdots i_N$.
Here we give the proof.
If $N=2$, it is obvious from the definition
that $\epsilon _{i_1i_2}$ is antisymmetric.
Let us assume that $\epsilon _{i_1\cdots i_N} \equiv \epsilon _{i_1i_2} \cdots \epsilon _{i_{N-1}i_N}$
is now antisymmetric and consider about the rank-$N+1$ tensor
\begin{equation}
\epsilon _{i_1\cdots i_{N+1}} \equiv \epsilon _{i_1\cdots i_N} \cdot
\epsilon _{i_1i_{N+1}} \cdots \epsilon _{i_{N}i_{N+1}}.
\end{equation}
We only need to prove that this tensor is antisymmetric for
indices $i_1$ and $i_{N+1}$.
One can rewrite the above form to
\begin{equation}
\epsilon _{i_1\cdots i_{N+1}} = \epsilon _{i_2\cdots i_N} \cdot
\epsilon _{i_1i_2} \cdots \epsilon _{i_1i_N} \cdot
\epsilon _{i_2i_{N+1}} \cdots \epsilon _{i_{N}i_{N+1}} \cdot
\epsilon _{i_1i_{N+1}}.
\end{equation}
When one exchanges the indices $i_1$ and $i_{N+1}$,
tensor $\epsilon _{i_2\cdots i_N}$ does not change.
Tensor $\epsilon _{i_1i_2} \cdots \epsilon _{i_1i_N}$
and $\epsilon _{i_2i_{N+1}} \cdots \epsilon _{i_{N}i_{N+1}}$
change into each other with the same factor $(-1)^{N-1}$, so their product does not change.
And finally, the tensor $\epsilon _{i_1i_{N+1}}$
changes its sign by $-1$.
Therefore, the tensor $\epsilon _{i_1\cdots i_{N+1}}$
changes its sign by $-1$ in this operation.
Thus the argument is proved.
It is trivial from the definition that 
the value of the tensor $\epsilon _{i_1\cdots i_N}$ takes only
$1$, $-1$ or $0$.\par
From this, one can conclude that 
the tensor $A_{i_1\cdots i_N}$ is antisymmetric.
We have proved that $\epsilon _{i_1\cdots i_N}$ is antisymmetric.
The vector product $c_{i_1}^i \cdots c_{i_N}^i$ is 
apparently symmetric.
Therefore, $c_{i_1}^i \cdots c_{i_N}^i \epsilon _{i_1\cdots i_N}$
is antisymmetric.
$A_{i_1\cdots i_N}$ is a linear combination of these antisymmetric tensors
for $i = 1,\cdots ,K$,
then antisymmetric.\par
Notice that the representation (\ref{expansion}) is an
approximative form of general antisymmetric tensor.
Antisymmetric tensor $A_{i_1\cdots i_N}$ has $_MC_N$ elements,
and belongs to the dimension $_MC_N$ vector space.
Therefore, if $K=_MC_N$ and all $c_{i_1}^i \cdots c_{i_N}^i \epsilon _{i_1\cdots i_N}$
are linearly independent for $i = 1,\cdots ,K$, 
then any $A_{i_1\cdots i_N}$ can be expanded in the form of (\ref{expansion}).
However, in the following section we will see that
one can well approximate at least some antisymmetric tensors 
which are solutions of the many body problem,
with the condition $K=1$.\par
Apparently, the vector product for $K=2$ is an extension of 
the vector product for $K=1$.
Generally, the vector product for $K+1$ is an extension of
the vector product for $K$.
Here we see that the vector product for $K=1$ includes 
all slater determinants for a given orbital set $\psi _1,\cdots ,\psi _M$.
When $K=1$, the vector product is given as
\begin{equation}
A_{i_1\cdots i_N} = c c_{i_1} \cdots c_{i_N} \epsilon _{i_1i_2} \cdots \epsilon _{i_{N-1}i_N}.
\end{equation}
We only need to consider the case which $i_1=1,\cdots ,i_N=N$.
When it is possible to expand the wave function $\psi (x_1\cdots x_N)$
in the orbital set $\psi _1,\cdots ,\psi _N$,
the antisymmetric tensor $A_{i_1\cdots i_N}$
has degree of $_NC_N=1$ and thus uniquely determined 
under the antisymmetric condition, with an
arbitrary factor.
The slater determinant for $\psi _1,\cdots ,\psi _N$
is of course antisymmetric.
Therefore, we only need to show that 
the vector product for $K=1$ can be antisymmetric
when expanded in orbitals $\psi _1,\cdots ,\psi _N$ and
zero when expanded in other orbitals.
This condition is satisfied 
when one takes the vector $c_{i_1}$ for
\begin{eqnarray}
c_{i_1} = \left\{ \begin{array}{ll}
1 & (i_1=1,\cdots ,N), \\
0 & (\mathrm{others}). \\
\end{array} \right.
\end{eqnarray}
Thus the argument is proved.\par

It is not difficult to see that 
the vector product with $K=1$ ($\psi _{VP}^1$) is indeed an extension
of the slater determinant
if the number of the orbital $M$ is larger than the
number of electrons $N$.
In other words, there are wave functions 
which are representable in the form of $\psi _{VP}^1$
but not in a slater determinant.
The characteristic of the slater determinant is that
the first order reduced density matrix ($\gamma _1$) of the slater determinant
has the same eigenvalues $1/N$  for $N$ natural orbitals.
One can easily see that $\gamma _1$ of an arbitrary wave function
which is representable as $\psi _{VP}^1$
has generally different eigenvalues $\lambda _i < 1/N$
by simply take random values for the vector $c_{i_1}$
and calculate $\gamma _1$ and its eigenvalues.
This fact indicates that $\psi _{VP}^1$
is generally interacting and 
not always representable by a single slater determinant.\par
As we discussed above, the vector product with $K=1$
is an extension of a single slater determinant.
For this reason, one can obtain the same or lower 
ground state energy $E_0$ 
by using $c_{i_1}$ as variational parameters,
compared to the Hartree-Fock approximation.
This is possible when one uses a orbital set $\psi _1,\cdots ,\psi _M$
which includes the Hartree-Fock derived orbitals $\psi _1,\cdots ,\psi _N$.
One can obtain further lower energy by the vector product method
when he takes variation also for orbitals $\psi _1,\cdots ,\psi _M$.
This variation of orbitals is possible in various way.
One way is to take unitary transform of these orbitals 
and take variation for the elements of the unitary matrix.
The details of this transform are to be explained in the following section.

\section{Energy calculation in the vector product method}

Generally, the expectation value of the energy $E$ of the $N$ electron system 
in the normalized state $\psi (x_1,\cdots ,x_N)$ is given as
\begin{equation}
E = \int \cdots \int dx_1\cdots dx_N \psi^*(x_1,\cdots ,x_N)H\psi(x_1,\cdots ,x_N).
\end{equation}
However, the hamiltonian of electrons is generally two body and in the following form:
\begin{equation}
H = \sum _{i=1}^N(-(1/2)\nabla _i^2+v(\vec{r}_i)) + \sum _{(i, j)}1/|\vec{r}_i-\vec{r}_j|,
\end{equation}
Therefore, one can obtain $E$ from the normalized second order reduced density matrix ($\gamma _2$)\cite{coleman}:
\begin{eqnarray}
E &=& N\int dx_1[(-(1/2)\nabla _1^2+v(\vec{r}_1))\gamma _1(x'_1,x_1)]_{x'_1=x_1}\nonumber \\
&+& N(N-1)/2\int \int dx_1dx_2 \frac{1}{|\vec{r}_1-\vec{r}_2|} \gamma _2(x_1x_2,x_1x_2),
\end{eqnarray}
\begin{equation}
\gamma _2(x'_1x'_2,x_1x_2)\equiv \int \cdots \int dx_3\cdots dx_N
\psi ^*(x'_1x'_2x_3\cdots x_N)\psi (x_1x_2x_3\cdots x_N),
\label{gamma2def}
\end{equation}
\begin{equation}
\gamma _1(x'_1,x_1)\equiv \int dx_2 \gamma _2(x'_1x_2,x_1x_2).
\end{equation}
One can obtain the tensor representation of $\gamma _2$ by
substituting (\ref{tensor}) for (\ref{gamma2def}):
\begin{equation}
\gamma _2(x_1x_2,x_3x_4) = \sum _{i_1i_2i_3i_4}C_{i_1i_2i_3i_4}
\psi _{i_1}^*(x_1)\psi _{i_2}^*(x_2)\psi _{i_3}(x_3)\psi _{i_4}(x_4),
\end{equation}
\begin{equation}
C_{i'_1i'_2i_1i_2} \equiv \sum _{i_3\cdots i_N}A^*_{i'_1i'_2i_3\cdots i_N}A_{i_1i_2i_3\cdots i_N}.
\label{gamma2}
\end{equation}
Here, orbitals $\psi _1,\cdots ,\psi _M$ are assumed to be orthonormal.
One can write down the tensor representation of $E$ as following:
\begin{equation}
E = \sum _{i_1i_2i_3i_4}h_{i_1i_2i_3i_4}C_{i_1i_2i_3i_4} / n,
\label{energy}
\end{equation}
\begin{eqnarray}
n&\equiv &\langle \psi (x_1\cdots x_N)|\psi (x_1\cdots x_N)\rangle \nonumber \\
&=& \sum _{i_1i_2}C_{i_1i_2i_1i_2},
\end{eqnarray}
\begin{equation}
h_{i_1i_2i_3i_4} \equiv N(t_{i_1i_3}+v_{i_1i_3})\delta _{i_2i_4} + \frac{N(N-1)}{2}w_{i_1i_2i_3i_4},
\end{equation}
\begin{equation}
t_{i_1i_3} \equiv \int dx_1 \psi ^*_{i_1}(x_1)(-1/2)\nabla ^2_1 \psi _{i_3}(x_1),
\end{equation}
\begin{equation}
v_{i_1i_3} \equiv \int dx_1 \psi ^*_{i_1}(x_1) v(\vec{r}_1) \psi _{i_3}(x_1),
\end{equation}
\begin{equation}
w_{i_1i_2i_3i_4} \equiv \int \int dx_1 dx_2 
\psi ^*_{i_1}(x_1)\psi ^*_{i_2}(x_2) \frac{1}{|\vec{r}_1-\vec{r}_2|} \psi _{i_3}(x_1) \psi _{i_4}(x_2).
\end{equation}\par
Next, we will see the representation of $\gamma _2$ 
in vector product.
Here we assume real values for the components of the vector product
$c_i$ and $c_{i1}^i$ for simplicity.
We need to substitute (\ref{expansion}) for (\ref{gamma2}).
The result is
\begin{eqnarray}
C_{i_1i_2i_3i_4}&=&\sum _{i,j=1}^{K}c_ic_j
c_{i_1}^ic_{i_2}^ic_{i_3}^jc_{i_4}^j
\epsilon _{i_1i_2}\epsilon _{i_3i_4} \nonumber \\
&\cdot &\sum _{j_3\cdots j_N}
c_{j_3}^ic_{j_3}^j\epsilon _{i_1j_3}\epsilon _{i_2j_3}\epsilon _{i_3j_3}\epsilon _{i_4j_3}
\cdots
c_{j_N}^ic_{j_N}^j\epsilon _{i_1j_N}\epsilon _{i_2j_N}\epsilon _{i_3j_N}\epsilon _{i_4j_N}
d_{j_3\cdots j_N} \nonumber \\
&\equiv & 
\sum _{i,j=1}^{K}c_ic_j
c_{i_1}^ic_{i_2}^ic_{i_3}^jc_{i_4}^j
\epsilon _{i_1i_2}\epsilon _{i_3i_4} \cdot I^{ij}_{i_1i_2i_3i_4},
\end{eqnarray}
\begin{eqnarray}
d_{j_3\cdots j_N} &\equiv &\epsilon _{j_3\cdots j_N}^2 \nonumber \\
&=& \left\{ \begin{array}{ll}
1 & (j_3\cdots j_N : \mathrm{all\, \,different}), \\
0 & (\mathrm{others}). \\
\end{array} \right.
\end{eqnarray}
There is a method to calculate the tensor $I^{ij}_{i_1i_2i_3i_4}$.
When one omits the indices $i_1i_2i_3i_4$ and $ij$,
$I^{ij}_{i_1i_2i_3i_4}$ has a following form:
\begin{equation}
I = \sum _{j_3\cdots j_N}a_{j_3}\cdots a_{j_N} d_{j_3\cdots j_N},
\label{dsum}
\end{equation}
\begin{equation}
a_{j_k} \equiv c_{j_k}^ic_{j_k}^j\epsilon _{i_1j_k}\epsilon _{i_2j_k}\epsilon _{i_3j_k}\epsilon _{i_4j_k}.
\end{equation}
From the definition of the tensor $d_{j_3\cdots j_N}$,
one can conclude that the sum in (\ref{dsum}) is 
taken over all permutation $(j_3\cdots j_N)$ for
$j_3\cdots j_N = 1,\cdots ,M$
with the condition that all $j_3\cdots j_N$ are different.
Therefore, the value of $I$ in (\ref{dsum}) can be 
represented as the $M-(N-2)$-th order coefficients of the following polynomial:
\begin{equation}
f(t) \equiv (N-2)!(t+a_1)\cdots (t+a_M).
\end{equation}
Then the problem is how to calculate the $M-(N-2)$-th order coefficients of
a given polynomial $f(t)$.
Notice that now $f(t)$ is an $M$-th order polynomial of $t$.
Then $f(t)$ can be expanded in a following form:
\begin{equation}
f(t) = b_0 + b_1t + \cdots + b_Mt^M,
\end{equation}
\begin{equation}
I = b_{M-(N-2)}.
\end{equation}
One can derive the $M-(N-2)$-th order coefficients $b_{M-(N-2)}$
by solving a linear algebra problem.
One can define a vector $\vec{b}$ of $M+1$ dimension from the 
unknown coefficients $b_0,\cdots ,b_M$.
One can have arbitrary $M+1$ different values for $t_0,\cdots ,t_M$.
Then he has
\begin{eqnarray}
f(t_0) &=& b_0 + b_1t_0 + \cdots + b_Mt_0^M, \nonumber \\
&\cdots &\nonumber \\
f(t_M) &=& b_0 + b_1t_M + \cdots + b_Mt_M^M.
\end{eqnarray}
One can solve this linear algebra problem 
with given values for $t_0,\cdots ,t_M$ and $f(t_0),\cdots ,f(t_M)$
and find all the coefficients $b_i$.
To do that, one only need to calculate the inverse matrix of 
a matrix $T_{ij} \equiv t_i^j$ and multiplies it to a vector
composed of $f(t_0),\cdots ,f(t_M)$.
In this way, one can calculate the value of $b_{M-(N-2)}$,
and therefore the value of tensor $I$ for each indices 
$i_1i_2i_3i_4$,$ij$.
Then one can calculate the matrix element $C_{i_1i_2i_3i_4}$
from $c_i$ and $c^i_{i1}$.\par
One has to take calculation steps proportional to $M^2$
for each calculation of $I^{ij}_{i_1i_2i_3i_4}$,
because he has to calculate the value of 
$M$-th order polynomial $f(t)$ for about $M$ times.
To calculate $C_{i_1i_2i_3i_4}$, 
one has to take this $M^2$ step for $M^4$ times for each indices,
therefore he needs calculation steps proportional to $M^6$ 
for each calculation of the matrix $C_{i_1i_2i_3i_4}$.

\section{Variation of orbitals}
In this section, we will consider about 
the variation of orbitals 
used in the expansion of many body wave function.
The result of the vector product method is depending
of the choice of orbitals,
then one has to take variation for orbitals
in the vector product method
in order to obtain lower energy.
Let us assume that we use orbitals $\psi _1,\cdots ,\psi _M$
and a basis set $\phi _1,\cdots ,\phi _B$.
Here, $B \geq M$ and orbitals are expanded in the basis set:
\begin{equation}
\psi _i(x) = \sum _{j=1}^BU_{ij}\phi _j(x).
\label{basis}
\end{equation}
We assume that the basis set is orthonormal.
Then, dimension-$B$ vectors $U_{1j},\cdots ,U_{Mj}$
should be orthonormal in order to maintain 
orbitals $\psi _1,\cdots ,\psi _M$ to be orthonormal.
Next, we will see how this expansion is applied 
on the expression of the energy matrix $h_{i_1i_2i_3i_4}$
appeared in (\ref{energy}).
The definition of $h_{i_1i_2i_3i_4}$ is
\begin{eqnarray}
h_{i_1i_2i_3i_4} &=& \int \int dx_1dx_2 
\psi _{i_1}(x_1)\psi _{i_2}(x_2)
[N(-(1/2)\nabla _1^2 + v(\vec{r}_1)) \nonumber \\
&+& \frac{N(N-1)}{2}\frac{1}{|\vec{r}_1-\vec{r}_2|}]
\psi _{i_3}(x_1)\psi _{i_4}(x_2).
\label{h}
\end{eqnarray}
By substituting (\ref{basis}) for (\ref{h}),
one gets following expression:
\begin{equation}
h_{i_1i_2i_3i_4} =
\sum _{j_1j_2j_3j_4=1}^B
U_{i_1j_1}U_{i_2j_2}U_{i_3j_3}U_{i_4j_4} H_{j_1j_2j_3j_4},
\end{equation}
\begin{eqnarray}
H_{j_1j_2j_3j_4} &\equiv &\int \int dx_1dx_2 
\phi _{j_1}(x_1)\phi _{j_2}(x_2)
[N(-(1/2)\nabla _1^2 + v(\vec{r}_1)) \nonumber \\
&+& \frac{N(N-1)}{2}\frac{1}{|\vec{r}_1-\vec{r}_2|}]
\phi _{j_3}(x_1)\phi _{j_4}(x_2).
\end{eqnarray}
This matrix $h_{i_1i_2i_3i_4}$ can be calculated by taking proportional to $MB^4$ steps
from the fixed matrix $H_{j_1j_2j_3j_4}$.
The variation of orbitals is possible when one takes variation for 
the matrix $U_{ij}$ with maintaining vectors $U_{1j},\cdots ,U_{Mj}$ to be orthonormal.
Notice that when $B=M$, $U_{ij}$ is a unitary matrix or orthogonal matrix.

\section{Results of calculation}
In this section, we will report on the results of 
the calculation of ground state energy
of carbon ($_6$C) and oxygen ($_8$O) atom,
using the vector product method with $K=1$.
The author of this paper admits that he is not
a specialist of computational physics.
He does not emphasize the numerical accuracy
of the results shown in this section.
The purpose of this section is to show 
that the calculation of the expectation energy
with the vector product is possible 
and one can obtain better results with the vector product
compared to the Hatree-Fock.
For calculations of the diagonalization in Hatree-Fock and Full CI,
we used CLAPACK
which is LAPACK usable in C.
\par
First, we will report on the results of $_6$C ($E_{VP}$)
by using the vector product method with $K=1$
with 14 atomic orbitals as a basis set,
which is 1s$^2$,$\cdots $, 3s$^2$ and 3p$_x^2$
with the effective nuclear charge $Z=5.5$.
In this calculation, we use $M=14$ orbitals
and set $B$ to 14.
We took variation for dimension-14 vector $c^1_{i_1}$ and 
a 14 $\times $ 14 orthogonal matrix $U_{ij}$.
The expectation value of the energy $E$ is 
given explicitly as a polynomial of 
$c^1_{i_1}$ and $U_{ij}$.
Therefore, we can explicitly define the differential value of $E$
for each $c^1_{i_1}$ and $U_{ij}$.
Then, we took variation for $c^1_{i_1}$ and $U_{ij}$
by using the steepest descent method.
We started from randomly chosen value for $c^1_{i_1}$
and a unit matrix for $U_{ij}$
as an initial value.
We also calculated the ground state energy in this basis set
by the Hartree-Fock method ($E_{HF}$) and Full CI method ($E_{FCI}$).
The results are following:
\begin{center}
\begin{tabular}[c]{|c|c|c|}
\hline
$M$=14 $_6$C&$E_{HF}$&$-37.661$ \\ \hline
$M$=14 $_6$C&$E_{VP}$&$-37.688$ \\ \hline
$M$=14 $_6$C&$E_{FCI}$&$-37.708$ \\ 
\hline
\end{tabular}
\end{center}
In this results, one can see that 
even though we set $K=1$ and used only one $\psi _{VP}^1$
as many body wave function,
he can obtain well better results
compared to the Hartree-Fock.
If one takes $K$ larger than 1, then 
it is sure that he will obtain further better results.
This result $E_{VP}=-37.688$ is obtained by
the steepest descent method and not yet converged.
Therefore we may rather say $E_{VP}\leq -37.688$.\par
Next, we will report on the results of $_6$C and $_8$O atom
using the vector product method with $K=1$ ($E_{VP}$)
with 19 atomic orbitals as a basis set,
which is 1s$^2$,$\cdots $, 3p$^6$ and 3d$_{3z^2-r^2}^1$.
We set the effective nuclear charge as
$Z=5.5$ for $_6$C and $Z=7.5$ for $_8$O.
We also calculated the ground state energy in this basis set
by the Hartree-Fock method ($E_{HF}$).
We also compare these results with results in a literature
$E_{HF}^G$, $E_{DFT}^G$ and $E_{EXP}^G$\cite{gunnarsson},
which are results in Hartree-Fock, Density functional theory
and experiment, respectively.
The results are following:
\begin{center}
\begin{tabular}[c]{|c|c|c|}
\hline
$M$=19 $_6$C&$E_{HF}$&$-37.666$ \\ \hline
$M$=19 $_6$C&$E_{VP}$&$-37.790$ \\ \hline
$_6$C&$E_{HF}^G$&$-37.702$ \\ \hline
$_6$C&$E_{DFT}^G$&$-37.479$ \\ \hline
$_6$C&$E_{EXP}^G$&$-37.858$ \\ 
\hline
\end{tabular}
\end{center}
\begin{center}
\begin{tabular}[c]{|c|c|c|}
\hline
$M$=19 $_8$O&$E_{HF}$&$-74.908$ \\ \hline
$M$=19 $_8$O&$E_{VP}$&$-74.956$ \\ \hline
$_8$O&$E_{HF}^G$&$-74.858$ \\ \hline
$_8$O&$E_{DFT}^G$&$-74.532$ \\ \hline
$_8$O&$E_{EXP}^G$&$-75.113$ \\ 
\hline
\end{tabular}
\end{center}
Here, $E_{VP}$ is not converged for both $_6$C
and $_8$O.
From these results, one can conclude that
the results of the vector product method with $K=1$
can be better than the results of Hatree-Fock.
If one takes $K$ larger than 1 then he will
obtain further lower results for $E_{VP}$.\par
In the calculation of the vector product method,
one needs proportional to $M^6$ steps for 
an evaluation of the matrix $\gamma _2$.
In $c^1_{i_1}$, There are $M$ variables for variation.
Therefore we took proportional to $M^7$ steps
for the variation of matrix $\gamma _2$.
In the variation of the orthogonal matrix $U_{ij}$,
one needs to take proportional to $M^5$ $(MB^4)$ steps
for each transformation of the energy matrix $h_{i_1i_2i_3i_4}$.
We took variation for each rows of the matrix.
Therefore we spend proportional to $M^6$ $(M^2B^4)$ steps 
for the variation of orbitals.
Then the estimated calculation time $T$ 
for the vector product method is 
\begin{equation}
T \sim O(M^7) + O(M^2B^4).
\end{equation}
The total amount of the calculation time 
for $E_{VP}$ was 
about a few hours for $_6$C and $_8$O ($M=19$)
with a $\sim 3$GHz CPU.
In our calculation, we spend larger amount of time
for the term $O(M^6)=O(M^2B^4)$ compared to $O(M^7)$.

\section{Conclusion}
We propose a new representation of many body electron wave function,
namely the vector product.
We also propose its application as a post Hatree-Fock method
to evaluate the ground state energy of many body electron systems.
The results of the vector product method will converge 
to the results of the Full CI method
when one takes sufficiently large value for the parameter $K$
and keeps each vector products $\psi _{VP}^i$ linearly independent.
We obtained systematically better energy results compared to the
results of the Hatree-Fock method
for $_6$C and $_8$O atoms.
The estimated calculation time $T$ 
for the vector product method is 
$T \sim O(M^7)$ as a function of orbital number $M$.
It is expected that using higher spec CPUs,
one can obtain the ground state energy better than Hartree-Fock
in more big systems.
In the vector product method, one can 
simultaneously variate the orbitals used in the calculation.
This is an advantage of the vector product method
compared to other methods such as CI method
in which orbitals are fixed during the calculation.
In the vector product method, 
one can simultaneously obtain the 
many body wave function of the system.
This means that one can obtain many
physical quantities of the system at the ground state.
For example, the off diagonal long range order (ODLRO) in solid
which is related to the superconductivity 
can be explained in a way that 
the maximum eigenvalue of the 
second order reduced density matrix ($\lambda _{max}^2$)
satisfies the following condition\cite{ueda}:
\begin{equation}
N^2\lambda _{max}^2 \sim O(N).
\end{equation}
Calculated wave functions 
and second order reduced density matrices in the vector product method
are not non-interacting in general.
Then there is a possibility that the superconductivity
of solid can be explained by the 
results of the vector product method.

\end{document}